\documentclass[twoside]{dis07}
\usepackage[latin1]{inputenc}
\usepackage[dvips]{graphicx,epsfig,color}
\usepackage{wrapfig,rotating}
\usepackage{amssymb,amsmath,array}

\begin{document}
\title{Deeply Virtual Compton Scattering at JLab Hall A}

\author{Eric Voutier\\
\it{fot the Jefferson Lab Hall A and DVCS Collaborations}
%
%
\vspace{.3cm}\\
%
LPSC, Universit\'e Joseph Fourier, CNRS/IN2P3, INPG \\
53 avenue des Martyrs, F-38026 Grenoble - France
%
}

\maketitle

\begin{abstract}
The deeply virtual Compton scattering reaction has been investigated in the 
Hall A of the Jefferson Laboratory by measuring longitudinally polarized 
$({\vec e},e' \gamma)$ cross sections, in the valence quark region, for protons 
and neutrons. In the proton channel, experimental results strongly support the 
factorization of the cross section at $Q^2$ as low as 2~GeV$^2$, opening the 
path to sytematic measurements of generalized parton distributions (GPDs). In 
the neutron case, preliminary data show sensitivity to the angular momentum of 
quarks~\cite{url}. 
\end{abstract}

\section{Introduction}

Over the ten past years, deeply virtual Compton scattering (DVCS) became the 
most promising process to explore the partonic structure of the
nucleon~\cite{{Die03},{Bel05}}. Similarly to the diffusion of light by a cristal, 
which tells about the internal structure and organization of the material, the 
scattering of energetic photon off the nucleon in the Bjorken regim ($Q^2>>M^2$ 
and $t<<Q^2$) allows to access the generalized parton distributions (GPDs) which 
describe the quark and gluon structure of the nucleon~\cite{{Ji98},{Col99}}. 
GPDs correspond to the coherence between quantum states of different (or same) 
helicity, longitudinal momentum, and transverse position and can be interpreted 
in the impact parameter space as a distribution in the transverse plane of 
partons carrying longitudinal momentum fraction $x
$~\cite{{Bur00},{Die02},{Bel02}}. The GPD framework provides a comprehensive 
picture of the nucleon structure which unifies within the same formalism form 
factors, structure functions, and partons angular momenta~\cite{Ji97}. 

In the Jefferson Laboratory (JLab) energy range, the Bethe-Heitler (BH) process, 
where the real photons are emitted either by the incoming or the scattered 
electrons, contributes significantly to the cross section of the 
electro-production of photons. However, the BH process is well-known and 
exactly calculable from the electromagnetic form factors of the nucleon. Then, 
similarly to holography technique, the BH process is used as a reference 
amplitude which interferes with the DVCS amplitude and magnifies the underlying 
effects~\cite{Ral02}. In JLab Hall A, two experimental observables have been 
investigated: the total $(e,e' \gamma)$ cross section
\begin{equation}
\frac{d^5 \sigma}{dQ^2 dx_B dt d\phi_e d \varphi} = {\cal T}_{BH}^2 + {\vert 
{\cal T}_{DVCS} \vert}^2 + 2 \, {\cal T}_{BH} \Re e\{ 
{\cal T}_{DVCS} \} \ ,
\label{eq:d5s}
\end{equation}
and the difference of polarized $({\vec e},e' \gamma)$ cross sections for 
opposite longitudinal beam helicities  
\begin{eqnarray}
\frac{d^5 \Sigma}{dQ^2 dx_B dt d\phi_e d \varphi} & = & \frac{1}{2} 
\left[ \frac{d^5 \overrightarrow{\sigma}}{dQ^2 dx_B dt d\phi_e d \varphi} - 
\frac{d^5 \overleftarrow{\sigma}}{dQ^2 dx_B dt d\phi_e d \varphi} \right]
\nonumber \\
& = & {\cal T}_{BH} \, \Im m \{{\cal T}_{DVCS}\} + \Re e \{{\cal T}_{DVCS}\} \,
\Im m \{{\cal T}_{DVCS}\} \ .
\label{eq:d5S}
\end{eqnarray}
While the former gives access to the real part of the DVCS amplitude, that is
the integral of a linear combination of GPDs convoluted with a quark 
propagator, the latter is a direct measurement of its imaginary part, which 
relates to a linear combination of GPDs in the handbag dominance
hypothesis~\cite{BelM02}. A dedicated experimental 
program~\cite{{E00-110},{E03-106}} was set to investigate the DVCS reaction off 
the proton and off the neutron, with the 
aim to test factorization in the proton channel and to explore the sensitivity 
of the neutron channel to $E_q$, the least known and constrained GPD. 

\section{Experimental apparatus}

A 5.75~GeV/c longitudinally polarized electron beam impinged on 15~cm liquid 
H$_2$ and D$_2$ cells, the latter serving as quasi-free neutron target. 
Scattered electrons were detected in the left High Resolution Spectrometer 
(HRS-L)~\cite{Alc04} for several $Q^2$ and constant $x_B$=0.36. Real photons were 
detected in a PbF$_2$ electromagnetic calorimeter organized in an 11$\times$12 
array of 3$\times$3$\times$18.6~cm$^3$ crystals centered around the direction of 
the virtual photon. The calorimeter front face was 110~cm from the target center 
supporting the useful $t$ acceptance -0.5~GeV$^2<t$. Typical beam intensities 
of 4~$\mu$A yielded a 4$\times$10$^{37}$~cm$^{-2} \cdot$s$^{-1}$ luminosity with 
76~\% polarized electrons. Three independent reactions were used to calibrate and monitor 
the calorimeter: H$(e,e'_{\mathrm{Calo.}}p_{\mathrm{HRS}})$, 
D$(e,e'_{\mathrm{Calo.}}\pi^-_{\mathrm{HRS}})pp$, and 
H,D$(e,e'_{\mathrm{HRS}}\pi^0_{\mathrm{Calo.}})X$~\cite{Maz06}. It should be emphasized 
that 
$\pi^-_{\mathrm{HRS}}$ and $\pi^0_{\mathrm{Calo.}}$ data are taken 
simultaneously with DVCS data, ensuring a continuous monitoring of the 
calibration and the resolution of the calorimeter.

\section{Factorization in p-DVCS}

\begin{wrapfigure}{rh}{0.50\columnwidth}
\centerline{\includegraphics[width=0.50\columnwidth]{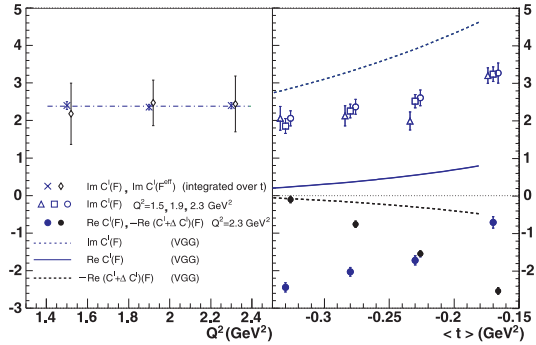}}
\caption{$Q^2$ and $t$ dependences of the GPDs linear combination extracted from
(un)polarized p-DVCS cross sections~\cite{Mun06}. The different curves (right panel) are
theoretical calculations from a GPD based model~\cite{Van99}.}\label{Fig:EV1}
\end{wrapfigure}
The polarized cross section difference (Eq.~\ref{eq:d5S}) for DVCS off the proton 
(p-DVCS) was measured at three different $Q^2$ ranging from 1.5~GeV$^2$ to 
2.3~GeV$^2$~\cite{Mun06}, and was analyzed according to the harmonic structure 
derived in Ref.~\cite{BelM02}. The $\sin(\phi)$ and $\sin(2 \phi)$ harmonic
coefficients (or moments) have been separated. In the context of this experiment, the kinematical factors 
entering the square of the DVCS amplitude suppress its contribution to $d^5 
\Sigma$ as compared to the BH$\cdot$DVCS interference amplitude, leading to a 
direct measurement of $\Im m \{{\cal T}_{DVCS}\}$. The $\sin(\phi)$ moment 
corresponds then to the imaginary part of the linear combination 
$C^I({\mathcal F})$ (Eq.~\ref{eq:tw2}) of the Compton form factors (CFFs) ${\mathcal H}$, 
${\widetilde {\mathcal H}}$, and ${\mathcal E}$ which relate to GPDs~\cite{BelM02}:
\begin{equation}
C^I({\mathcal F}) = F_1 {\mathcal H} + \xi (F_1+F_2) {\widetilde {\mathcal
H}} - \frac{t}{4M^2} F_2 {\mathcal E} \ . \label{eq:tw2}
\end{equation}
Figure~\ref{Fig:EV1} shows the $Q^2$ dependence of the twist-2 (Eq.~\ref{eq:tw2}) and 
twist-3 ($\Im m [C^I({\mathcal F}^{eff})]$) harmonic coefficients of  
$d^5\Sigma$: the observed independence on $Q^2$ is an indication for 
factorization. Furthermore, the contribution of the twist-3 terms to 
$d^5\Sigma$ was found to be small~\cite{Mun06}. These features are a strong 
indication that factorization applies even at $Q^2$ as low as 2~GeV$^2$.  

\section{Importance of the DVCS amplitude}

\begin{wrapfigure}{rh}{0.45\columnwidth}
\centerline{\includegraphics[width=0.45\columnwidth, height=125pt]{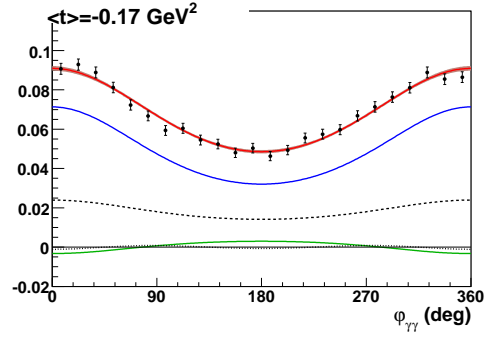}}
\caption{The $\phi$-dependence of the $d^4 \sigma$ differential cross section (Eq.~\ref{eq:d5s}
integrated over $\phi_e$) in nb/GeV$^4$ at $Q^2$=2.3~GeV$^2$, decomposed in BH and DVCS 
contributions~\cite{Mun06}.}
\label{Fig:EV2}
\end{wrapfigure}
The unpolarized H$(e,e' \gamma)$p cross section was also measured at the highest 
$Q^2$ point. Neglecting the DVCS$\cdot$DVCS term, the 
real part of the DVCS amplitude (Eq.~\ref{eq:d5s}) was extracted according to
the harmonic structure of Ref.~\cite{BelM02}. This leads to a $\cos(\phi)$ 
and $\cos(2 \phi)$ dependence, the gluon contribution - which would appear as 
a $\cos(3 \phi)$ term - being negligible in the valence quark region. 
Experimental data (points) are shown on Fig.~\ref{Fig:EV2} as a function of 
$\phi$ for the smallest $|t|$-bin. The red curve fitting the data is the sum of 
the different contributions to the cross section: deviations from the pure BH
amplitude (blue solid curve) shows that the DVCS amplitude contributes 
significantly to $d^5 \sigma$. This feature suggests that one should pay attention 
to the $\phi$-dependence of the denominator when extracting GPDs from beam spin 
asymetries. \\
In addition to the real part of the CFFs combination of Eq.~\ref{eq:tw2}, the extracted 
harmonic coefficients give access to the combination
\begin{equation}
C^I({\mathcal F})+\Delta C^I({\mathcal F}) = F_1 {\mathcal H} - \frac{t}{4M^2} 
F_2 {\mathcal E} - \xi^2 (F_1+F_2) ({\mathcal H} + {\mathcal E})
\end{equation}
which is independent of ${\widetilde {\mathcal H}}$. As for $d^5 \Sigma$, the contribution 
of twist-3 terms to $d^5 \sigma$ was found negligible, supporting again 
factorization~\cite{Mun06}. 

\begin{wrapfigure}{rh}{0.515\columnwidth}
\centerline{\includegraphics[width=0.515\columnwidth, height=106pt]{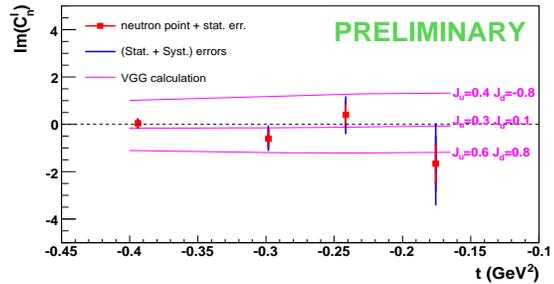}}
\caption{$t$-dependence of the $\sin(\phi)$ moments of the n-DVCS reaction~\cite{Maz06}. The
different curves correspond to GPD based calculations for different values of the $u$
and $d$ quarks contributions to the nucleon spin.}
\label{Fig:EV3}
\end{wrapfigure}

\section{Hunting quark angular momentum with n-DVCS}

Measuring the DVCS polarized cross section difference on a neutron target (n-DVCS), 
one can access, similarly to the proton, the combination of Eq.~\ref{eq:tw2}. 
Because of the smallness of the Dirac form factor and the cancellation between
the polarized $u$ and $d$ quark distributions in $\widetilde{\mathcal H}$, 
Eq.~\ref{eq:tw2} is dominated by the ${\mathcal E}$ contribution. This spin-flip 
GPD, which cannot be constrained by deep inclusive scattering, is of particular 
importance in Ji's sum rule leading to the quark angular momentum~\cite{Ji97}. 
The n-DVCS cross section difference $d^5\Sigma$ was deduced from the subtraction 
of hydrogen data to deuterium data at $Q^2$=1.9~GeV$^2$ and
$x_B$=0.36~\cite{Maz07-1}. The remaining coherent (d-DVCS) and incoherent (n-DVCS) 
contributions were extracted taking advantage of their $\Delta M_X^2$=$-t/2$ kinematical 
separation~\cite{Maz07-2} in the reconstructed squared missing mass, and the twist-2 (Eq.~\ref{eq:tw2}) harmonic coefficient
was obtained for several $t$ values, neglecting the higher twist contributions as
supported by p-DVCS data. Figure~\ref{Fig:EV3}~\cite{Maz06} shows the $t$-dependence of
the $\sin(\phi)$ moments extracted for the n-DVCS channel. They appear to be globally 
compatible with zero. The comparison to GPD based model calculations~\cite{Van99} shows 
the sensitivity of the present data to the contribution of the $u$ and $d$ quarks to the
nucleon spin.

\section{Conclusions}

The DVCS experimental program at JLab Hall A delivered its first results: the
factorization of the cross section was observed, and the power of neutron targets 
to reach quark angular momenta was proven. These features open unambiguously the 
era of systematic measurements of generalized parton distributions in DVCS 
processes at JLab 6 GeV, and 12 GeV in a near future.

\section*{Acknowledgments}

This work was supported in part by DOE contract DOE-AC05-06OR23177 under which 
the Jefferson Science Associates, LLC, operates the Thomas Jefferson National 
Accelerator Facility, the National Science Foundation, the French Atomic Energy 
Commission and the National Center for Scientific Research.


\begin{footnotesize}



%

\end{footnotesize}


\end{document}